%% file: Rayleigh_Channel_Binary_Diversity.tex
\documentclass[conference]{IEEEtran}
\ifCLASSINFOpdf
\else
\fi

\usepackage[T1]{fontenc}
\usepackage{graphics}
\usepackage{graphicx}
\usepackage{multirow}
\usepackage{amssymb}
\usepackage{textcomp}

\begin{document}
\title{Binary Diversity for Non-Binary LDPC Codes over the Rayleigh Channel}

\author{\IEEEauthorblockN{Matteo Gorgoglione$^*$, Valentin Savin$^*$, David Declercq$^\#$}
\IEEEauthorblockA{$^*$CEA-LETI, Minatec Campus, Grenoble, France \{matteo.gorgoglione, valentin.savin\}@cea.fr\\
$^\#$ETIS, ENSEA Univ. Cergy-Pontoise/CNRS, Cergy-Pontoise, France, declercq@ensea.fr}}

\maketitle

\begin{abstract}
In this paper we analyze the performance of several bit-interleaving strategies applied to Non-Binary Low-Density Parity-Check (LDPC) codes over the Rayleigh fading channel.
The technique of bit-interleaving used over fading channel introduces diversity which could provide important gains in terms of frame error probability and detection.

This paper demonstrates the importance of the way of implementing the bit-interleaving, and proposes a design of an optimized bit-interleaver inspired from the Progressive Edge Growth algorithm. This optimization algorithm depends on the topological structure of a given LDPC code and can also be applied to any degree distribution and code realization.

In particular, we focus on non-binary LDPC codes based on graph with constant symbol-node connection $d_v = 2$. These regular $(2,dc)$-NB-LDPC codes demonstrate best performance, thanks to their large girths and improved decoding thresholds growing with the order of Finite Field. Simulations show excellent results of the proposed interleaving technique compared to the random interleaver as well as to the system without interleaver.

\end{abstract}

\section{Introduction}
Since their rediscovery by MacKay \cite{MacKay} in 1996, Low-Density Parity-Check codes have attracted a lot of attention because they exhibit rates close to the Shannon capacity \cite{RichardsonUrbanke} for many transmission channels, despite their low decoding complexity.
With the evolution of the technology, new families of LDPC codes defined on non-binary alphabets have been proposed and studied. They demonstrate better performance with respect to the binary case, especially for moderate code lengths \cite{DaveyMacKay} but at the expense of more complex decoding architectures.

Non-binary LDPC codes can be defined by considering a non-binary alphabet $\mathcal{A}$, which for practical reasons is often considered to be endowed with a vector-space structure over the binary field $\mathbb{F}_2 = \{ 0, 1 \}$, and
a semigroup $\mathcal{G}$ acting on $\mathcal{A}$. A non-binary code of length $N$ is hence defined as the set of solutions $\mathbf{s} \in \mathcal{A}$ of a linear system $\mathbf{H} \mathbf{s}^T = \mathbf{0}$, where $\mathbf{H}$ is a matrix with coefficients in $\mathcal{G}$, referred to as the parity-check matrix of the code.

LDPC codes are decoded using the belief propagation (BP) algorithm  based on an iterative exchange of messages between nodes \cite{Gallager,Declercq}.

The case of codes for which the underlying bipartite graph is ultra-sparse, in the sense that each symbol-node is connected
to exactly $d_v = 2$ linear constraint-nodes, is of particular interest.
First, very large girths can be obtained for Tanner graphs with $d_v = 2$, as demonstrated in \cite{Xiao}, \cite{Venkiah}.
It has also been pointed out \cite{Sassatelli}, \cite{DaveyMacKay} that when the size of the non-binary alphabet grows, best decoding thresholds are obtained for average density of edges closer and closer to $d_v = 2$.
Practically, for NB-LDPC codes defined over Galois fields $\mathbb{F}_q$ with $q \geq 64$, best codes, both asymptotically and at finite lengths, are \textit{ultra-sparse} codes.
Despite those advantages, the \textit{ultra-sparse} LDPC codes in $\mathbb{F}_q$ suffer from a serious drawback, as their minimum distance is limited and grows at best as $\mathcal{O}(log(N))$ \cite{Poulliat}. This limitation is however not critical when the desired  error rate is above $10^{-5}$, which is the case of the wireless transmissions that we target in this paper.

\begin{figure}[!t]
\centering
\includegraphics[height=0.87\linewidth,angle=-90]{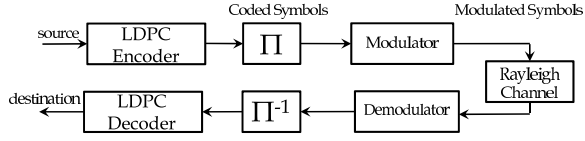}\\
\vspace{0.3cm}
\caption{\small{Transmission Chain}}
\label{fig:2.01}
\end{figure}

Radio channels in multipath environments, such as mobile or indoor contexts, can be modeled by a Rayleigh distributed fading \cite{Ozarow}. A classical way to fight against the fading effects of Rayleigh model is to introduce binary diversity by the mean of a bit-interleaver at the transmitter side. Such interleaver that operates between the encoder and the symbol mapper (see \figurename~\ref{fig:2.01}) drastically improves error rates in most of situations involving a fading channel.

In this paper, we demonstrate that in the case we make use of NB-LDPC codes for the forward error correction, the bit-interleaver is still of great importance to reach good performance, and moreover, we show
that the performance are even better when the bit-interleaver is well fitted to the NB-LDPC code structure. The proposed optimized interleaving algorithm is inspired from the Progressive Edge Growth algorithm, and associates the consecutive channel bits to the most \textit{separated} symbol-nodes.

The rest of the paper is organized as follows. A brief introduction of the NB-LDPC codes and the channel model are discussed in Section~\ref{sect1}. In Section~\ref{sect2} the interleaving algorithm is presented. Performance analysis are shown in Section~\ref{sect3}. Finally, Section~\ref{sect4} concludes the paper.



\section {Background and Notation} \label{sect1}

\subsection {NB-LDPC codes over $\mathbb{F}_q$}

We consider a non-binary alphabet $\mathbb{F}_q$ with $q = 2^p$ elements. We fix once for all the isomorphism:
$$\mathbb{F}_q \; \tilde{ \longrightarrow } \; \mathbb{F}_2^p $$
Elements of $\mathbb{F}_q$ are called \textit{symbols}, and their images under the above isomorphism are called \textit{binary images}. A non-binary LDPC code over $\mathbb{F}_q$ is defined as the kernel of a sparse matrix $\textbf{H} \in \textbf{M}_{M \times N} (\mathbb{F}_q)$. Thus, each codeword $\mathbf{s}$ composed of $N$ symbols, represents the solution of the linear system $\textbf{H} \mathbf{s}^T = \textbf{0}$.

Associated with the matrix \textbf{H} is the Tanner Graph \cite{Tanner}, which is constituted by N symbol-nodes on the top, representing the coded symbols, and M constraint-nodes on the bottom, representing, with the labeled edges, the linear constraints between these symbols (see \figurename~\ref{fig:2.02}). Each incident edge from \textit{symbol-node} $s_i$ to \textit{constraint-node} $c_j$ corresponds to a non-zero entry $h_{j,i} \in \textbf{H}$.

We define \textit{node degree} as the number of incident edges into such a node. In particular, $d_v$ indicates the variable node degree, whereas $d_{c}$ indicates the constraint-node degree.
In \figurename~\ref{fig:2.02} a Tanner graph for a (2,4)-regular code is represented in which all the symbol-nodes are of degree $d_v = 2$ and all the constraint-nodes are of degree $d_c = 4$.
\subsection{Modulation and Channel Model}
In this paper, we assume a Rayleigh channel model, typical of a  mobile/multipath environment. Therefore, the envelope of the channel response follows this distribution:
$$ p(h) = \frac{h}{\sigma^2} \exp ^ {- \frac{h^2}{2 \sigma^2}}$$
and the received symbol $\rho_k$ is:
$$\nonumber	\rho_k = h_k x_k + z_k$$
where $x_k$ stands for the $k$th transmitted modulated symbol scaled by i.i.d. Rayleigh factors $h_k$, and $z_k$ is the Additive White Gaussian Noise with variance $\sigma^2$.

The symbols $x_k$ are modulated by a $\mathcal{M}$-QAM constellation in which each mapped symbol is associated to $m=\log_2{\mathcal{M}}$ bits; we denote by $N_m$ the number of modulated symbols. 
\subsection{Belief Propagation Decoding}
The Belief Propagation (BP) decoder \cite{Wiberg} is based on an iterative \textit{message-passing} algorithm. For non-binary LDPC codes, the \textit{extrinsic} messages that circulate on the graph are multidimensional messages ({\it i.e.} vectors).

The initialization messages are Likelihood probability weights,  dependent on the channel statistics. For NB-LDPC codes, the decoder input consists of $N$ Likelihood vectors $\left( P (s_i = a) \right)_{a \in \mathcal{A}}$, where $s_i$ denotes the $i$th transmitted symbol, $i = \{1, \dots, N\}$.

In the case the order of the constellation and the order of the coded symbols match, the Likelihood vectors are directly derived from the demapped symbols. On the contrary, when a de-interleaver is placed before the decoder, the demapper performs a marginalization to transform the symbol likelihoods into bit-wise likelihoods, inducing
a performance loss due to the marginalization. Nevertheless, the
loss of information due to marginalization is counterbalanced by the
gain that the bit-interleaver brings in the case of fading channels.
We show in the result section that the diversity gain surpasses greatly the loss due to marginalization both in the waterfall and in the error floor regions.

Now let us present in short the non-binary decoder equations.
Let $P(v=a)$ denotes the probability that a random variable $v \in \mathbb{F}_q$ takes on value $a$. A message exchanged between a symbol-node and a check-node is a probability vector of size $q$:
$$\textbf{P}(v)= [ P(v=0),\dots, P(v=q-1) ]$$
of a random variable $v \in \mathbb{F}_q$.

Let $\alpha (i \rightarrow j)$ be the message from symbol-node $s_i$ to an adjacent check-node $c_j$ and $\beta (j \rightarrow i)$ be the message from a check-node $c_j$ to an adjacent symbol-node $s_i$. With $\gamma_i$ we indicate the channel likelihood for the symbol-node $s_i$.

Consider $\mathbf{s} \in \mathcal{C}$ a transmitted codeword of an LDPC code $\mathcal{C}$. The decoder aims to detect a codeword $\widehat{\mathbf{s}} \equiv \mathbf{s}$. The decoding process is generally composed by two half-iterations: i) update of symbol-node messages $\alpha(i \rightarrow j)$, taking into account the check-node messages $\beta(i \rightarrow j')_{j' \neq j}$ and the channel realizations $\gamma_i$, ii) update of check-node messages $\beta(j \rightarrow i)$ taking into account the symbol-node messages $\alpha(i' \rightarrow j)_{i' \neq i}$. At the end of each iteration, the decoder computes the symbol probabilities relative to each symbol-node in order to make a decision about $\widehat{\mathbf{s}}$.

Usually, one makes the decoder stop in two situations: {\it i)} a maximum number of iterations is reached, then $\widehat{\mathbf{s}}$ is computed from the messages at the last iteration, or ii) the syndrome is verified $\mathbf{H}\widehat{\mathbf{s}}^T=\mathbf{0}$, and a codeword $\widehat{\mathbf{s}}$ is identified.

A \textit{detected} error happens if $\widehat{\mathbf{s}}$ does not belong to the codeword set. If $\widehat{\mathbf{s}}$ belongs to the codeword set but it is not the transmitted codeword, the decoder makes an \textit{undetected} error. Undetected errors are due to codewords with a low Hamming weight, which is one the weaknesses of the considered $(2,d_c)$-NB-LDPC codes. We will make in section \ref{sect3} a detailed study of the percentages of detected and un-detected errors in each interleaving situation.
\begin{figure}[!t]
\centering
\includegraphics[height=0.8\linewidth,angle=-90]{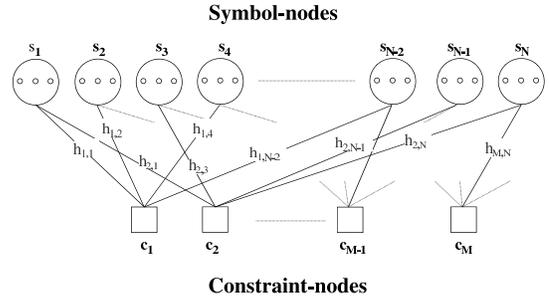}\\
\vspace{0.3cm}
\caption{\small{Non-Binary LDPC codes - Tanner Graph}}
\label{fig:2.02}
\end{figure}

Finally, we discuss the computation of the input Likelihood messages in each of the considered cases. Let $\Xi$ be the vector of received symbols and $\Gamma$ be the Likelihood messages at the inputs of symbol-nodes. In the case of $N_m = N$ and no bit-interleaver, there exists a one-to-one correspondence between the modulated and coded symbols:
$$
\begin{array}{c}
\Xi = [ \; {\rho}_{x_1}, \dots {\rho}_{x_k}, \dots {\rho}_{x_{N_m}} \;]\\
\;\; \downarrow \; \; \; \; \; \; \; \; \; \; \downarrow  \; \; \; \; \; \; \; \; \; \;\downarrow \;\\
\Gamma = [ \; {\gamma}_{s_1},\; \dots {\gamma}_{s_i}, \dots {\gamma}_{s_{N}} \;]\;\\
\end{array}
$$

The other case corresponds to the use of a bit-interleaver, or when the
size of the constellation does not match the size of the coded symbols ($N_m \neq N$). In such a case, an intermediate operation is used to transform the Likelihoods between the bits of the mapped symbols and the bits of the coded symbols:
$$
\begin{array}{c}
{\Xi} = [ \;{\rho}_{x_1}, \dots {\rho}_{x_{N_m}} \; ]\\
\; \; \; \; \; \; \; \; \; \; \; \; \; \; \; \Downarrow  \; \; \; \; \; \; \; \\
\Xi_{bin} = [ ( {\rho}_{x_{1,1}}, \dots {\rho}_{x_{1,m}} ) \dots ( {\rho}_{x_{N_m,1}}, \dots {\rho}_{x_{N_{m,m}}} ) ]\\
\; \; \; \; \; \; \; \; \; \; \; \; \; \; \; \Downarrow  \; \; \; \; \; \; \; \\
\Gamma = [ \; {\gamma}_{{s_1}},\; \dots {\gamma}_{s_{N}} \;]\;\\
\end{array}
$$
where $\Xi_{bin}$ contains the $n_m$ received bits. These bits are grouped in vectors of $p$ bits in order to obtain the messages which compose $\Gamma$. In the rest of the paper, we consider only the case where $m = p$, so that the latter marginalization transformation is used only when a bit-interleaver is employed.


\section {Interleaving algorithm}\label{sect2}

The effect of a bit-interleaver is to spread the coded bits in different modulation symbols, such that these bits composing a single coded symbol are affected by different fading factors. The advantage of using an interleaver is that the deep fading effects on the received bits are mitigated by the fact that those bits are distributed among the codewords after the de-interleaver.

The bit-interleaver can be seen as the construction of a superimposed regular $(m,p)$-bipartite graph (from now on called simply \textit{interleaving graph}) on the Tanner Graph. It connects the \textit{modulation-nodes} $x_k$ to the symbol-nodes $s_i$ of a pre-designed Tanner Graph. The modulation-nodes are another type of nodes representing the modulated symbols in the interleaving graph.

\begin{figure}[!t]
\includegraphics[height= \linewidth ,angle=270]{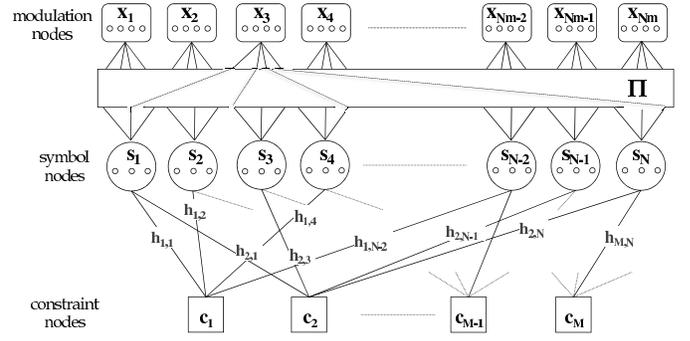}\\
\caption{\small{Global graph}}
\label{fig:2.03}
\end{figure}

In the interleaving graph, edges connect $N_m$ modulation symbol-nodes $x_k$ to $N$ coded symbol-nodes $s_i$ (see \figurename~\ref{fig:2.03}). In this part of the graph, $d_k$ denotes the modulation symbol degree, while $d_i$ denotes the coded symbol degree. To simplify the study, we can assume that each modulation symbol has constant modulation-node degree $d_k = m$ and that each coded symbol has constant interleaving symbol-node degree $d_i = p$.

Of course, since the number of coded bits $n$ can be computed either as the number of bits within the $N$ coded-symbols $s_i$, or as the number of bits within the $N_m$ modulated-symbols $x_k$, we have that $n = N p = N_m m$.

\figurename~\ref{fig:2.03} shows an example of a superimposed interleaving graph on the Tanner graph of \figurename~\ref{fig:2.02}. The interleaver is represented as a block denoted as $\mathbf{\Pi}$, with $m=4$ edges for each modulation-node and $p=3$ edges for each  coded symbol-node.

Therefore, the aim of the a bit-interleaving design is to look for an \textit{interleaving pattern} that contains the scrambled bit positions in the interleaving graph (upper part of the graph in the \figurename~\ref{fig:2.03}).

Though random interleaving shows already good performance as demonstrated in the section \ref{sect3}, we have devised an algorithm to optimize the interleaver design, which even more improves the performance, especially in the error-floor region. The bit-interleaver optimization algorithm is presented in the next section.

\subsection{PEG Optimization Interleaving}

Our bit-interleaver design is inspired from the Progressive Edge-Growth (PEG) method \cite{Xiao} used for constructing Tanner graphs with large girths that progressively connects symbols and constraint-nodes. For this reason, from now on, our optimization algorithm will be identified as PEG interleaving algorithm.

The PEG interleaving algorithm is efficient for creating good connections between the modulation and the symbol-nodes in a best effort way. Good connections are meant to give the largest possible girth to both the LDPC Tanner graph and the interleaving graph. Starting with the knowledge of $m$, $p$ and the LDPC Tanner graph topology, the algorithm connects each modulation-node to $m$ symbol-nodes. The rationale behind the optimization algorithm is to find, for each modulation-node, the $m$ most distant coded symbol-nodes (from a topological distance point of view), and therefore to build connections in the interleaving graph which results in the best girth. It should also be noted that the bit-interleaver design is code-dependent. As a matter of fact, the girth computation during the interleaver design takes into account the topology of the already designed NB-LDPC code $\mathcal{C}$. It results in particular that the bit-interleaver built with our algorithm is actually {\it matched} to a particular NB-LDPC code, which explains the further performance gains that we observe.\\

We now explain the principles of the PEG interleaving algorithm.
Recall that $d_{i}$ denotes the coded symbol-node degree and $d_{k}$ denotes the modulation-node degree. Before the algorithm starts, all the degrees are set to $0$. During the algorithm execution, the current node degrees represent the number of established connections for the nodes. Thus, for a coded symbol-node $s_i$, $i \in \{1, \dots, N\}$, the degree range is $0 \leq d_{i} \leq p$; for a modulation-node $x_k$, $k \in \{1, \dots, N_m\}$, the degree range is $0 \leq d_{k} \leq m$.

Let $x_k$ be a modulation-node to be connected. The PEG interleaving algorithm chooses the first connection between $x_k$ and $s_w$. $s_w$ is a randomly chosen symbol-node among the available ones with the lowest current degree. Then, we expand both the Tanner and the interleaving graphs through the three types of nodes taking into account the new connection. Once the graph expansion is complete, the bottom of this graph is the set of symbol-nodes $\{s_z\}$ that are the most distant symbol-nodes from $s_w$. Hence, the algorithm can connect the modulation-node $x_k$ to one of the coded symbol-node $s_z \in \{s_z\}$ with the lowest degree. A new connection is thereby chosen, and the algorithm goes to the next edge selection, by performing the same steps - graph expansion and node selection. The procedure is iterated until it reaches the correct degree $d_{k} = m$. The algorithm stops when all the modulation-nodes are connected.

Note that with this PEG interleaving algorithm, only the coded symbol index is important for the global girth of the graph, and not the location of the bit {\it inside} the coded symbol binary map. Actually, an extra local scrambler could be added --- at the coded symbol level --- without impacting the girth of the global graph. In our simulations, however, this extra local scrambler did not impact significantly on the error rate performance.

Although the above described construction could be adapted to any NB-LDPC code, including codes with irregular node distributions, we restricted our study to regular (2,$d_c$)-LDPC codes. For these codes, the error rate simulation results and the study of detected vs. undetected errors are conducted in the next section.

\begin{figure}[!t]
\centering
\includegraphics[width=\linewidth]{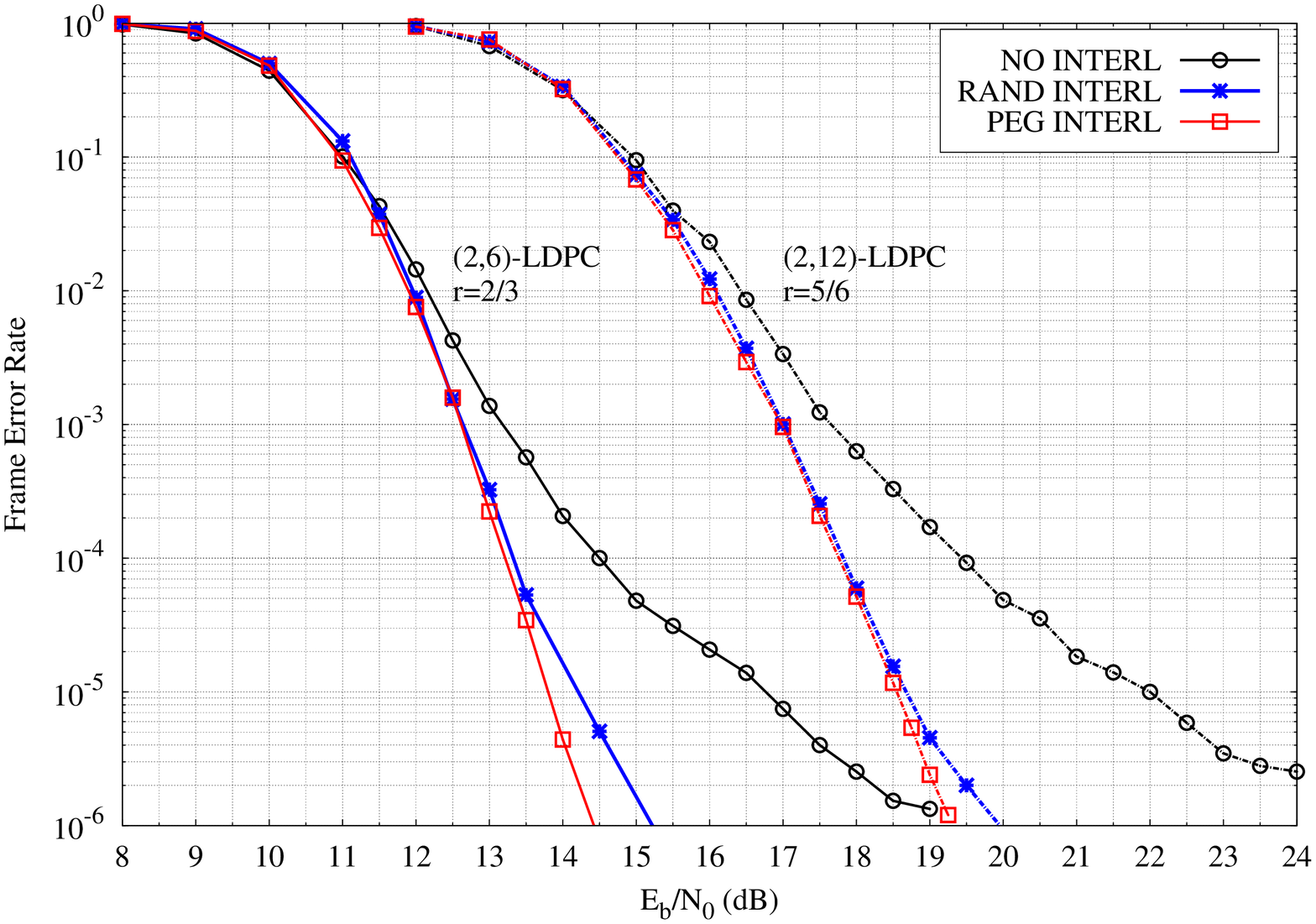}\\
\caption{\small{Frame Error Rates for (2,6) and (2,12) NB-LDPC codes, $\;\;n = 612$, $\mathbb{F}_{64}$, $64$-QAM}}
\label{fig:3.01}
\end{figure}
\begin{figure}[!t]
\centering
\includegraphics[width=\linewidth]{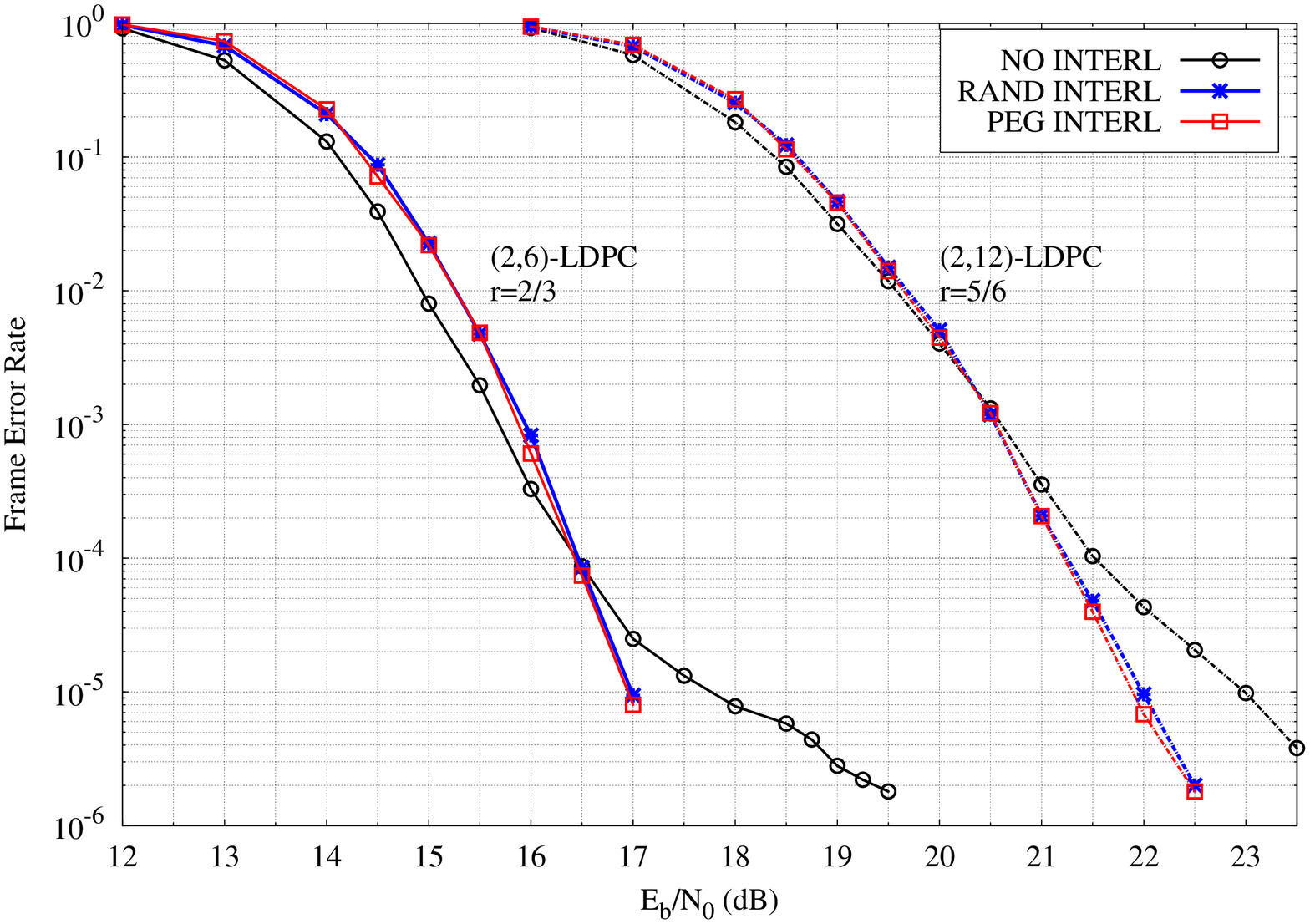}
\caption{\small{Frame Error Rates for (2,6) and (2,12) NB-LDPC codes, $\;\;n = 816$, $\mathbb{F}_{256}$, $256$-QAM}}
\label{fig:3.02}
\end{figure}

\section{Simulation Results} \label{sect3}
In this section, we present the simulation results for the bit-interleaved and not bit-interleaved NB-LDPC codes. Motivated by the good asymptotical thresholds that grows with the order field, we have focused on LDPC code alphabets $\mathbb{F}_{64}$ and $\mathbb{F}_{256}$. It has been shown that, for those high orders, the ultra-sparse NB-LDPC codes with node degree $d_v = 2$  is the best choice. We have therefore simulated LDPC codes for those two different alphabets and two different coding rates: {\it i)} (2,6)-NB-LDPC codes corresponding to a rate $R = 2/3$ and {\it ii)} (2,12)-NB-LDPC codes corresponding to a rate $R = 5/6$. These NB-LDPC code Tanner graphs are designed with PEG method \cite{Xiao} (minimum girth $g = 6$) and decoded by using a Belief-propagation decoder \cite{Declercq}. As said in the previous section, the size of the alphabet is the same of the size of modulation constellation: codes defined over $\mathbb{F}_{64}$ are transmitted using the $64$-QAM modulation, whereas codes defined over $\mathbb{F}_{256}$ are transmitted using the $256$-QAM modulation.

Each LDPC code is simulated on the Rayleigh fading channel with three possible transmission systems: first, a non-interleaved system with direct mapping from the modulation to the coded symbols, then with a random bit-interleaver and finally with our PEG-optimized interleaver.\\

The curves in \figurename~\ref{fig:3.01} represent the Frame Error Rates (FER) of the considered (2,$d_c$)-LDPC codes defined over $\mathbb{F}_{64}$ for small codewords ($n = 612$ bits, $N = 102$ coded symbols) modulated with a $64$-QAM. A significant performance gain can be observed both in the waterfall and in the error floor regions    in the  presence of a bit-interleaver. This shows that the diversity gain brought by the bit-interleaving surpasses greatly the information loss due to symbol-to-bit marginalization.

As expected, the PEG interleaved system (represented by $\square$)  achieves roughly the same gain as random interleaving (represented by $\ast$) in the waterfall region, but shows also an improved error-floor. This was our goal of building bit-interleavers that optimize the girth of the global graph.

\begin{figure}[!t]
\centering
\includegraphics[width=\linewidth]{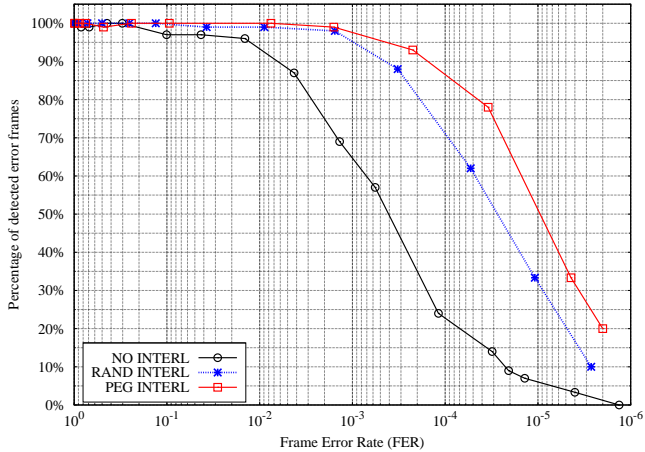}
\caption{\small{Percentage of detected frame errors for a (2,6)-regular NB-LDPC code, $n = 612$, $\mathbb{F}_{{64}}$, $64$-QAM}}
\label{fig:3.03}
\end{figure}

\begin{figure}[!t]
\centering
\includegraphics[width=\linewidth]{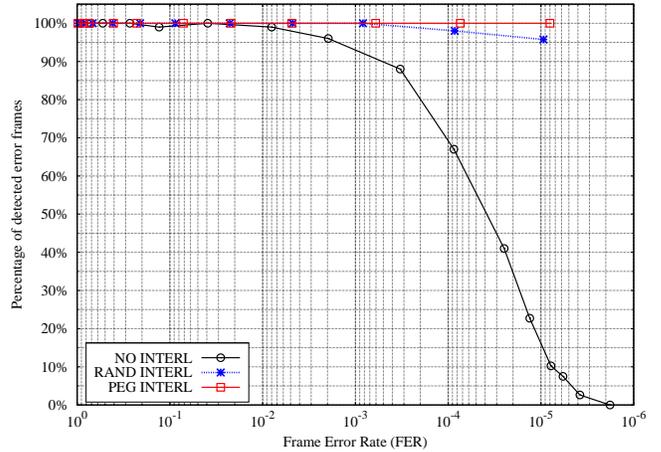}
\caption{\small{Percentage of detected error frames for a (2,6)-regular NB-LDPC code, $n = 612$, $\mathbb{F}_{{256}}$, $256$-QAM}}
\label{fig:3.04}
\end{figure}

In \figurename~\ref{fig:3.02} we show the performance of the codes defined over $\mathbb{F}_{256}$ with codelength $n=816$ ($N=102$ coded symbols) and modulated with a $256$-QAM. As can be seen, when the size of alphabet grows, bit-interleavers still gain in the error floor region, but the gap between the two kinds of interleavers vanishes.
Now, let us discuss the effect of bit-interleaving and our optimized construction for the detection of frame errors. We have drawn in
\figurename~\ref{fig:3.03} and ~\ref{fig:3.04} the percentage of detected frame errors with respect to the FER, for the (2,6)-LDPC codes defined respectively in $\mathbb{F}_{64}$ and in $\mathbb{F}_{256}$.

Although it was expected, this study confirms that the better performance in the error floor region results from a better detection of frame errors. More precisely, bit-interleaving on a Rayleigh channel helps the decoder to avoid convergence to low-weight codewords, therefore improving at the same time the performance and the probability of error detection. This last feature is very interesting since in most wireless mobile transmissions, a link adaptation strategy implementing retransmission of detected wrong frames is generally used (ARQ or Hybrid-ARQ).

Note that even in the case of codes in $\mathbb{F}_{256}$, our bit-interleaving optimization shows an interesting gain in detected frame errors ($100\%$ for all FER simulated) compared to the random-interleavers, although the error rates were the same (see \figurename~\ref{fig:3.02}).


\section{Conclusions}\label{sect4}

In this paper we investigate the non-binary LDPC codes transmitted over a Rayleigh fading channel. Since modulated symbols can be affected by different fading factors, deep fading could make some codeword symbols totally unrecoverable in case of one-to-one correspondence between modulated and coded symbols, leading to a poor system performance.

In order to avoid this phenomenon, binary diversity can be exploited by using a bit-interleaver module placed between the encoder and the modulator. A random interleaver and an optimized interleaver have been analyzed by running simulations over short size regular $(2,d_c)$-LDPC codes. 

The	 optimized interleaving algorithm is inspired from the Progressive Edge Growth (PEG) method and it ensures maximum girth of the global graph. Although the bit demapping needed in the interleaved leads to information loss, it has been demonstrated that the use of the interleaves ensures improved frame-error probabilities compared to a system without it.
Additionally, in all considered cases, the optimized interleaver showed an even better gain with respect to the random interleaver, as far as performance and error detection rates are concerned.

\bibliographystyle{IEEEfull.bib}

\input{./Bibliography}

\end{document}

%% file: Bibliography.tex



%% file: Rayleigh_Channel_Binary_Diversity.bbl
\addcontentsline{toc}{chapter}{Bibliography}

\begin{thebibliography}{100}

\bibitem{MacKay} D. J. C. MacKay and R. M. Neal, "Near Shannon limit performance of low density parity check codes," \textit{Electronics Letters}, vol.32, no.18, pp. 1645-1646, August 1996, Reprinted \textit{Electronics Letters}, vol 33, no. 6, March 1997, 457-458.

\bibitem{RichardsonUrbanke} T. Richardson, M.A. Shokrollahi and R. Urbanke, "Design of Capacity-Approaching irregular Low-Density Parity-Check codes", \textit{IEEE Trans. Inform. Theory}, vol 47, no. 2, pp. 619-637, Feb. 2001.

\bibitem{DaveyMacKay} M. Davey and D. J. C. MacKay, "Low Density Parity Check Codes over GF(q)", \textit{IEEE Commun. Lett.}, vol. 2, pp. 165-167, June 1998.

\bibitem{Gallager} R. G. Gallager, "Low-density Parity-Check Codes", \textit{IRE Transaction on Information Theory}.

\bibitem{Declercq} D. Declercq and M. Fossorier, "Decoding algorithms fon nonbinary LDPC codes over GF(q)," \textit{IEEE Trans. on Comm.}, vol.55, no. 4, April 2007.

\bibitem{Xiao} X.-Y. Hu, E. Eleftheriou, and D. M. Arnold, \textit{Regular and irregular progressive edge-growth tanner graph}, \textit{IEEE Trans. Inform. Theory}, vol 51, no. 1, pp. 386-398, Jan. 2005.

\bibitem{Venkiah} A. Venkiah, D. Declercq and C. Poulliat, "Design of cages with a randomized progressive edge growth algorithm," \textit{IEEE Commun. Letters}, vol. 12, no. 4, pp. 301-303, April 2008.

\bibitem{Sassatelli} L. Sassatelli and D. Declercq, "Non-binary hybrid LDPC codes," \textit{in IEEE Trans. Info. Theo.}, vol. 56, no. 10, pp. 1069-1074, Dec. 2005.

\bibitem {Poulliat} C. Poulliat, M. Fossorier  and D. Declercq, "Design of regular (2, $d_c$)-LDPC codes over GF(q) using binary images", \textit{IEEE Trans. Commun.}, vol.56, no. 10, pp. 1626-1635, October 2008.

\bibitem{Ozarow} L. H. Ozarow, S. Shamai and A. D. Wyner, "Information theoretic considerations for cellular mobile radio," \textit{IEEE Trans. on Vehicular Tech.}, vol. 43, no. 2, pp. 359-378, May 1994.

\bibitem{Tanner}R. M. Tanner, "A recursive approach to low complexity codes", \textit{IEEE Trans. Inf. Theory}, vol. IT-27, no. 9, pp. 533-548, Sep. 1981.

\bibitem{Wiberg} N. Wiberg, "Codes and decoding on general graphs," Ph.D. dissertation, Linköping Univ., Linköping, Sweden, 1996.






\end{thebibliography}
